\documentclass[aps,prl,twocolumn,groupedaddress]{revtex4-1}
\usepackage[]{graphicx}
\usepackage[english]{babel}
\usepackage[latin1]{inputenc}
\usepackage{natbib}
\usepackage{bm}
\usepackage{color}
\usepackage{ulem}
\begin{document}

% Use the \preprint command to place your local institutional report
% number in the upper righthand corner of the title page in preprint mode.
% Multiple \preprint commands are allowed.
% Use the 'preprintnumbers' class option to override journal defaults
  % to display numbers if necessary
%\preprint{}

%Title of paper
\title{Highly Effective Superconducting Vortex Pinning in Conformal Crystals}

% repeat the \author .. \affiliation  etc. as needed
% \email, \thanks, \homepage, \altaffiliation all apply to the current
% author. Explanatory text should go in the []'s, actual e-mail
% address or url should go in the {}'s for \email and \homepage.
% Please use the appropriate macro foreach each type of information

% \affiliation command applies to all authors since the last
% \affiliation command. The \affiliation command should follow the
% other information
% \affiliation can be followed by \email, \homepage, \thanks as well.
\author{S. Gu\'enon}
\email[]{stefan.guenon@physics.ucsd.edu}
\affiliation{Department of Physics and Center for Advanced Nanoscience,
  University of California-San Diego, La Jolla, California 92093, USA}
\author{Y. J. Rosen}
\affiliation{Department of Physics and Center for Advanced Nanoscience, University of California-San Diego, La Jolla, California 92093, USA}
\author{Ali C. Basaran}
\affiliation{Department of Physics and Center for Advanced Nanoscience, University of California-San Diego, La Jolla, California 92093, USA}
\author{Ivan K. Schuller}
\affiliation{Department of Physics and Center for Advanced Nanoscience, University of California-San Diego, La Jolla, California 92093, USA}
%\email[]{Your e-mail address}
%\homepage[]{Your web page}
%\thanks{}
%\altaffiliation{}
%Collaboration name if desired (requires use of superscriptaddress
%option in \documentclass). \noaffiliation is required (may also be
%used with the \author command).
%\collaboration can be followed by \email, \homepage, \thanks as well.
%\collaboration{}
%\noaffiliation

\date{\today}
\begin{abstract}
%%% Introduction
%
We have investigated the vortex dynamics in superconducting thin film devices
with non-uniform patterns of artificial pinning centers
(APCs). 
The magneto-transport properties of a conformal crystal and a randomly diluted APC
pattern are compared with that of a triangular
reference lattice.
We have found that in both cases the magneto-resistance below the first matching field of
the triangular reference lattice is significantly reduced. 
For the conformal crystal, the magneto-resistance is below the
noise floor indicating highly effective vortex pinning over a
wide magnetic field range. 
Further, we have discovered that for asymmetric patterns the R vs. H curves
are mostly symmetric.
This implies that the enhanced vortex pinning is due to the commensurability
with a stripe in the non-uniform APC pattern and not due to a rearrangement
and compression of the whole
vortex lattice.
(submitted on 04/29/13 to APL)
\end{abstract}
% insert suggested PACS numbers in braces on next line
% insert suggested keywords - APS authors don't need to do this
%\keywords{(submitted on 04/29/13 to APL)}

%\maketitle must follow title, authors, abstract, \pacs, and \keywords
\maketitle

%%% INTRODUCTION 
% 
Vortex dynamics in superconducting thin films with artificial pinning centers (APCs)
have been extensively studied in recent years.
A central problem in this field is how to improve the critical
current density of a superconducting device by choosing a suitable APC distribution.
If the APCs are arranged in a hexagonal or a rectangular lattice, the
critical current is increased for certain magnetic matching fields due to the 
commensurability with the Abrikosov vortex lattice \cite{Baert1995,
Harada1996, Martin1997, Castellanos1997, Moshchalkov1998, Morgan1998, Villegas2003}.  
In order to increase the field range for vortex pinning, different APC
distributions like quasiperiodic \cite{Kemmler2006, Misko2010} or
random \cite{Rosen2010, Kemmler2009} lattices have been investigated.
Recent theoretical papers are focusing on non-uniform APC distributions like
hyperbolic-tessellation arrays \cite{Misko2012} and
conformal crystals \cite{Ray2012}.
A conformal crystal can be obtained by conformally mapping a semiannular section
of a regular hexagonal lattice on a rectangle (see \cite{Ray2012,Rothen1993,Rothen1996}). 
In this transformation the angles are preserved, but a vertex 
density gradient along one side of the rectangle is introduced.
Hence, a conformal crystal is a non-uniform pattern, in which on a small scale the vertices
are arranged triangularly. 
On a large scale however, the vertex density changes
continuously.\\ 
A recent theoretical study \cite{Ray2012} has shown that the pinning in conformal crystals
is significant stronger over a much wider field range than that found for other APCs with an equivalent number of pinning sites.
These promising results have not yet been confirmed experimentally.\\
In our experimental study we have investigated the vortex dynamics in superconducting thin
film devices with non-uniform APC patterns.
We compare the magneto-transport properties of conformal crystals and
corresponding randomly diluted APC patterns with the properties of a triangular
reference lattice.
Additionally, we investigate whether an electric current applied to a superconducting
micro bridge with a non-uniform APC pattern produces enhanced pinning due to
vortex distribution rearrangement and compression.    
Because the current pushes the vortices against the edge of the sample, an
asymmetric APC pattern with respect to the center line of the
bridge could produce an asymmetric R vs. H curve when compared to the
  triangular array. 
In agreement with theoretical calculations, we find enhanced pinning by
non-uniform APC patterns. 
However, contrary to the expectations the R vs. H
curves are essentially symmetric in all cases indicating 
that the enhanced vortex pinning is not the result of a large scale
redistribution of vortices.
It appears rather that the vortices are effectively pinned locally in a
stripe, in which the vortex lattice and the non-uniform APC pattern is commensurate.\\
\begin{figure}
\includegraphics{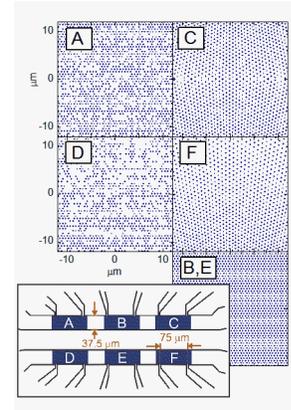}
\caption{(color online) Artificial pinning center patterns in the
  center of the micro bridge devices. A: symmetric randomly diluted, B and E:
  triangular, C: symmetric conformal crystal, D: asymmetric randomly diluted,
  F: asymmetric conformal crystal. Inset: Geometry of the sample under
  investigation. The micro bridge devices A-C and D-F are in series.}
\label{fig:pattern}
\end{figure}
Standard e-beam writing was used to prepare magnetic pinning sites \cite{Velez2008}.
First, the APC dot patterns were written into PMMA resist on a SiO$_2$ coated
Si substrate. 
Then, a 40 nm thick Cobalt film was deposited via e-beam evaporation and
approximately 120 nm diameter Cobalt dots were obtained in a lift-off step. 
Micro bridges of a 100 nm thick, RF-sputtered Nb film were prepared using optical lithography and lift-off.
A superconducting transition temperature of 8.6 K was measured.\\
Fig. \ref{fig:pattern} shows the centers of the APC patterns and the geometry of
the sample. 
To avoid misalignment of the APC pattern with respect to the
superconducting micro bridges, they were all written in a single scan with
an approximate write field of $400 \times 400\,\mu\textnormal{m}^2$. 
The APC patterns are $100\,\mu$m long and $40\,\mu$m wide, to assure that the
whole area between the voltage taps is filled with APCs. 
Device A and C are symmetric (the pinning site density increases to both edges
of the micro bridge), while D and F are asymmetric (the pinning site
density increases to the lower edge of the micro bridge). 
The triangular lattices B and E have a lattice constant of approximately
$0.6\,\mu$m.
The conformal crystals (C and F) were obtained by mapping the triangular lattice according
to \cite{Ray2012}, but, instead of choosing a semiannular region, a
partial annular section was chosen with an opening angle of $\pi/2$.
Hence, in this study the conformal crystal is deformed less than in \cite{Ray2012}.
For the randomly diluted lattices (A and D) the vertices of the triangular
lattice were randomly removed, so that the pinning site density is the same as
that of the corresponding conformal crystals.
All patterns were compressed by 18\% in the direction perpendicular to the micro bridge edges.\\
Magneto-resistance was measured in a liquid Helium cryostat
with a superconducting magnet and a variable temperature insert. 
The sample was cooled with evaporated Helium gas and a
temperature stability better than $1\,$mK was achieved using a heater in the sample mount.
There was a temperature drift of a few mK on a time scale of a few minutes due
to a change in the evaporation rate and therefore the time slot for acquiring an R vs. H
was about one minute.\\
In this study, it was very important to eliminate the different thermo-electric
off-sets in the wiring. 
Therefore, we used for the symmetric patterns the DC polarity reversed mode
of the Keithley transport electronics \cite{Daire2005}.
In this mode the current source changes polarity in a low frequency square
wave pattern and the voltage is read out synchronously.
The signal is averaged by subtracting the opposite polarity voltages.
For the asymmetric pattern we used an alternative technique to differentiate between
the two current directions.  
We acquired four R vs. H (DC) curves by changing the current
source polarity for consecutive curves. 
In this technique the average resistivity is given by
\[\overline{R}(H)=\left[R_1(H)-R_2(-H)+R_3(H)-R_4(-H)\right]/4\]
The minus sign before the magnetic field in the second and fourth
R(H) is necessary, because a change in the current direction changes the
direction of the Lorentz-force.\\  
\begin{figure}
\includegraphics{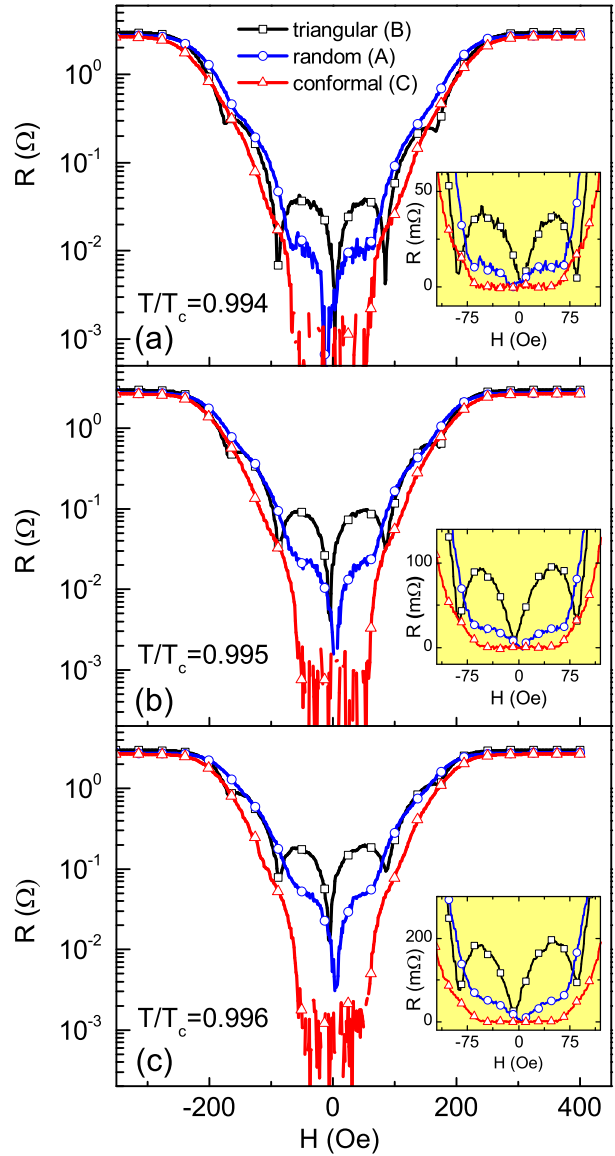}
\caption{(color online) Resistance vs. magnetic field of the micro bridge
  devices with symmetric APC patterns at different set temperatures (logarithmic
  scale). The ratio $T/T_c$ was estimated for the triangular reference
  lattice. Insets: Resistance vs. magnetic fields for small fields (linear scale).}
\label{fig:symmetric}
\end{figure}
In order to investigate the flux flow resistance of non-uniform APC patterns we have
measured the magneto-resistance at different temperatures near the
superconducting transition.  
In all electrical transport measurements presented, we applied a relatively
small $10\,\mu$A current to avoid Joule heating.
With the series arrangement, we were able to consecutively measure the R vs. H
curves of the three devices at each temperature by changing the voltage contacts. 
We used the micro bridge devices with a triangular APC pattern as a
reference. 
The ratio $T/T_c$ was determined from the magneto-resistance vs. temperature dependence
of the triangular reference pattern.\\
Fig. \ref{fig:symmetric} shows R vs. H curves of the devices with symmetric patterns at three different
temperatures. 
The triangular reference lattices shows two matching minima at
approximately 86 Oe and 172 Oe due to the relatively small diameter to
separation ratio of the magnetic pinning dots \cite{Hoffmann2000}.
The resistance of the device with a randomly diluted APC is considerably smaller than that of the triangular lattice below the first matching field. 
Above the first matching field, R vs. H is almost the same except in
vicinity of the second matching field where the triangular lattice is smaller.
The flux flow resistance of the device with the conformal crystal is smaller
than that of the randomly diluted or the triangular APC pattern for all
temperatures and magnetic fields except in vicinity of the matching fields (fig. \ref{fig:symmetric} a for $T/T_c=0.994$).
For field values smaller than the first matching field, the flux flow
resistance is below the noise floor of the measurement.
This indicates strong pinning which immobilizes the vortices.
We emphasize that this behavior is reproducible and robust for all
temperatures in the superconducting state (see fig. \ref{fig:symmetric}).\\
\begin{figure}
\includegraphics{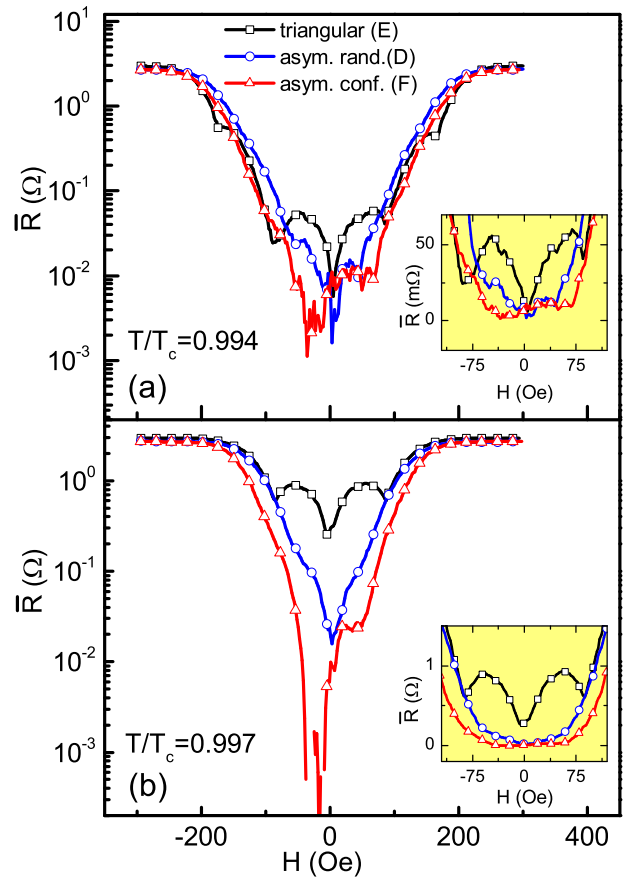}
\caption{(color online) Averaged resistance vs. magnetic field of the micro bridge
  devices with asymmetric APC patterns at different set temperatures (logarithmic
  scale). The ratio $T/T_c$ was estimated for the triangular reference
  lattice. Insets: Averaged resistance vs. magnetic fields for small fields (linear scale).}
\label{fig:asymmetric}
\end{figure}
The R vs. H of the asymmetric devices are very similar to the symmetric ones as
shown in fig. \ref{fig:asymmetric}.
For resistances above approximately $30\,\textnormal{m}\Omega$, the R vs. H of the
asymmetric conformal crystal and the randomly diluted pattern are symmetric. 
For small resistance values the R vs. H curves are slightly asymmetric.
For instance, consider the asymmetric conformal crystal 
at $\textnormal{T}/\textnormal{T}_\textnormal{c}=0.994$ (see fig \ref{fig:asymmetric} a).
The difference between the magneto-resistance at field values of -30\,Oe and
30\,Oe is approximately 6.7\,m$\Omega$. 
But in this set of graphs the R vs. H of the triangular reference lattice is
asymmetric, too.
Between the magneto-resistance at the first matching field the difference is
approximately 15\,m$\Omega$.
Hence, the R vs. H asymmetry of the symmetric reference pattern is more
pronounced.  
This indicates that the R vs. H asymmetry of the asymmetric pattern is smaller
than the systematic error.\\
In conclusion, we discovered that below the first matching field of the
triangular reference lattice, the R vs. H of the randomly diluted and the
conformal crystal is reduced.
For the conformal crystal, the pinning effectively reduces the resistance below
the noise floor over a wide magnetic field range.\\
In order to investigate whether the enhanced pinning is due to a
  compression and rearrangement of the vortex lattice, we have measured the 
magneto-transport of an asymmetric conformal crystal.
For a fixed current direction, the vortices are pushed towards the side with
either a high or low APC density when the field direction is changed. 
If the vortex distribution is rearranged and compressed by the Lorentz force
when it is pushed against the Bean-Livingston barrier at the
micro bridge edge then the commensurability of the vortices with the APC
distribution would depend on the field direction. 
This would result in an asymmetric R vs. H curve.   
However, our samples do not show a pronounced asymmetry when compared to the
triangular lattice.
We therefore suggest an alternative explanation:\\ 
For every magnetic field smaller than the first matching field there exists a
stripe in the non-uniform APC distribution, in which the APC density is more
or less equal to the vortex density.
The commensurability of the vortex lattice with the pinning sites in this
stripe, which is parallel to the micro bridge edge, increases the vortex
pinning and hinders or suppresses the movement of the whole vortex lattice. 
Although the conformal crystal and the randomly diluted lattice have the same
APC density, the commensurability with the vortex lattice
is considerably higher in the first one.
Therefore, the pinning in the conformal crystal is more effective. 
This study might be helpful to enhance the critical current density in tape
conductors, to reduce the 1/f-noise in SQUIDs \cite{Woerdenweber2000} or to
improve the performance of microwave resonators \cite{Bothner2011} in magnetic fields.
\begin{acknowledgments}
We thank D. Ray, C.J. Olson Reichhardt, B. Janko and C. Reichhardt for sharing their preprint and useful conversations. 
This work was supported by the Office of Basic Energy Science, U.S. Department of Energy, under Grant No. DE FG03-87ER-45332. 
\end{acknowledgments}
% Create the reference section using BibTeX:
%merlin.mbs apsrev4-1.bst 2010-07-25 4.21a (PWD, AO, DPC) hacked
%Control: key (0)
%Control: author (8) initials jnrlst
%Control: editor formatted (1) identically to author
%Control: production of article title (-1) disabled
%Control: page (0) single
%Control: year (1) truncated
%Control: production of eprint (0) enabled
%


\begin{thebibliography}{20}%
\makeatletter
\providecommand \@ifxundefined [1]{%
 \@ifx{#1\undefined}
}%
\providecommand \@ifnum [1]{%
 \ifnum #1\expandafter \@firstoftwo
 \else \expandafter \@secondoftwo
 \fi
}%
\providecommand \@ifx [1]{%
 \ifx #1\expandafter \@firstoftwo
 \else \expandafter \@secondoftwo
 \fi
}%
\providecommand \natexlab [1]{#1}%
\providecommand \enquote  [1]{``#1''}%
\providecommand \bibnamefont  [1]{#1}%
\providecommand \bibfnamefont [1]{#1}%
\providecommand \citenamefont [1]{#1}%
\providecommand \href@noop [0]{\@secondoftwo}%
\providecommand \href [0]{\begingroup \@sanitize@url \@href}%
\providecommand \@href[1]{\@@startlink{#1}\@@href}%
\providecommand \@@href[1]{\endgroup#1\@@endlink}%
\providecommand \@sanitize@url [0]{\catcode `\\12\catcode `\$12\catcode
  `\&12\catcode `\#12\catcode `\^12\catcode `\_12\catcode `\%12\relax}%
\providecommand \@@startlink[1]{}%
\providecommand \@@endlink[0]{}%
\providecommand \url  [0]{\begingroup\@sanitize@url \@url }%
\providecommand \@url [1]{\endgroup\@href {#1}{\urlprefix }}%
\providecommand \urlprefix  [0]{URL }%
\providecommand \Eprint [0]{\href }%
\providecommand \doibase [0]{http://dx.doi.org/}%
\providecommand \selectlanguage [0]{\@gobble}%
\providecommand \bibinfo  [0]{\@secondoftwo}%
\providecommand \bibfield  [0]{\@secondoftwo}%
\providecommand \translation [1]{[#1]}%
\providecommand \BibitemOpen [0]{}%
\providecommand \bibitemStop [0]{}%
\providecommand \bibitemNoStop [0]{.\EOS\space}%
\providecommand \EOS [0]{\spacefactor3000\relax}%
\providecommand \BibitemShut  [1]{\csname bibitem#1\endcsname}%
\let\auto@bib@innerbib\@empty
%</preamble>
\bibitem [{\citenamefont {Baert}\ \emph {et~al.}(1995)\citenamefont {Baert},
  \citenamefont {Metlushko}, \citenamefont {Jonckheere}, \citenamefont
  {Moshchalkov},\ and\ \citenamefont {Bruynseraede}}]{Baert1995}%
  \BibitemOpen
  \bibfield  {author} {\bibinfo {author} {\bibfnamefont {M.}~\bibnamefont
  {Baert}}, \bibinfo {author} {\bibfnamefont {V.~V.}\ \bibnamefont
  {Metlushko}}, \bibinfo {author} {\bibfnamefont {R.}~\bibnamefont
  {Jonckheere}}, \bibinfo {author} {\bibfnamefont {V.~V.}\ \bibnamefont
  {Moshchalkov}}, \ and\ \bibinfo {author} {\bibfnamefont {Y.}~\bibnamefont
  {Bruynseraede}},\ }\href {\doibase 10.1103/PhysRevLett.74.3269} {\bibfield
  {journal} {\bibinfo  {journal} {Phys. Rev. Lett.}\ }\textbf {\bibinfo
  {volume} {74}},\ \bibinfo {pages} {3269} (\bibinfo {year}
  {1995})}\BibitemShut {NoStop}%
\bibitem [{\citenamefont {Harada}\ \emph {et~al.}(1996)\citenamefont {Harada},
  \citenamefont {Kamimura}, \citenamefont {Kasai}, \citenamefont {Matsuda},
  \citenamefont {Tonomura},\ and\ \citenamefont {Moshchalkov}}]{Harada1996}%
  \BibitemOpen
  \bibfield  {author} {\bibinfo {author} {\bibfnamefont {K.}~\bibnamefont
  {Harada}}, \bibinfo {author} {\bibfnamefont {O.}~\bibnamefont {Kamimura}},
  \bibinfo {author} {\bibfnamefont {H.}~\bibnamefont {Kasai}}, \bibinfo
  {author} {\bibfnamefont {T.}~\bibnamefont {Matsuda}}, \bibinfo {author}
  {\bibfnamefont {A.}~\bibnamefont {Tonomura}}, \ and\ \bibinfo {author}
  {\bibfnamefont {V.~V.}\ \bibnamefont {Moshchalkov}},\ }\href {\doibase
  10.1126/science.274.5290.1167} {\bibfield  {journal} {\bibinfo  {journal}
  {Science}\ }\textbf {\bibinfo {volume} {274}},\ \bibinfo {pages} {1167}
  (\bibinfo {year} {1996})}\BibitemShut {NoStop}%
\bibitem [{\citenamefont {Mart\'in}\ \emph {et~al.}(1997)\citenamefont
  {Mart\'in}, \citenamefont {V\'elez}, \citenamefont {Nogu\'es},\ and\
  \citenamefont {Schuller}}]{Martin1997}%
  \BibitemOpen
  \bibfield  {author} {\bibinfo {author} {\bibfnamefont {J.~I.}\ \bibnamefont
  {Mart\'in}}, \bibinfo {author} {\bibfnamefont {M.}~\bibnamefont {V\'elez}},
  \bibinfo {author} {\bibfnamefont {J.}~\bibnamefont {Nogu\'es}}, \ and\
  \bibinfo {author} {\bibfnamefont {{\relax Ivan K}.}~\bibnamefont
  {Schuller}},\ }\href {\doibase 10.1103/PhysRevLett.79.1929} {\bibfield
  {journal} {\bibinfo  {journal} {Phys. Rev. Lett.}\ }\textbf {\bibinfo
  {volume} {79}},\ \bibinfo {pages} {1929} (\bibinfo {year}
  {1997})}\BibitemShut {NoStop}%
\bibitem [{\citenamefont {Castellanos}\ \emph {et~al.}(1997)\citenamefont
  {Castellanos}, \citenamefont {Wordenweber}, \citenamefont {Ockenfuss},
  \citenamefont {v.d. Hart},\ and\ \citenamefont {Keck}}]{Castellanos1997}%
  \BibitemOpen
  \bibfield  {author} {\bibinfo {author} {\bibfnamefont {A.}~\bibnamefont
  {Castellanos}}, \bibinfo {author} {\bibfnamefont {R.}~\bibnamefont
  {Wordenweber}}, \bibinfo {author} {\bibfnamefont {G.}~\bibnamefont
  {Ockenfuss}}, \bibinfo {author} {\bibfnamefont {A.}~\bibnamefont {v.d.
  Hart}}, \ and\ \bibinfo {author} {\bibfnamefont {K.}~\bibnamefont {Keck}},\
  }\href {\doibase 10.1063/1.119701} {\bibfield  {journal} {\bibinfo  {journal}
  {Appl. Phys. Lett.}\ }\textbf {\bibinfo {volume} {71}},\ \bibinfo {pages}
  {962} (\bibinfo {year} {1997})}\BibitemShut {NoStop}%
\bibitem [{\citenamefont {Moshchalkov}\ \emph {et~al.}(1998)\citenamefont
  {Moshchalkov}, \citenamefont {Baert}, \citenamefont {Metlushko},
  \citenamefont {Rosseel}, \citenamefont {Van~Bael}, \citenamefont {Temst},
  \citenamefont {Bruynseraede},\ and\ \citenamefont
  {Jonckheere}}]{Moshchalkov1998}%
  \BibitemOpen
  \bibfield  {author} {\bibinfo {author} {\bibfnamefont {V.~V.}\ \bibnamefont
  {Moshchalkov}}, \bibinfo {author} {\bibfnamefont {M.}~\bibnamefont {Baert}},
  \bibinfo {author} {\bibfnamefont {V.~V.}\ \bibnamefont {Metlushko}}, \bibinfo
  {author} {\bibfnamefont {E.}~\bibnamefont {Rosseel}}, \bibinfo {author}
  {\bibfnamefont {M.~J.}\ \bibnamefont {Van~Bael}}, \bibinfo {author}
  {\bibfnamefont {K.}~\bibnamefont {Temst}}, \bibinfo {author} {\bibfnamefont
  {Y.}~\bibnamefont {Bruynseraede}}, \ and\ \bibinfo {author} {\bibfnamefont
  {R.}~\bibnamefont {Jonckheere}},\ }\href {\doibase 10.1103/PhysRevB.57.3615}
  {\bibfield  {journal} {\bibinfo  {journal} {Phys. Rev. B}\ }\textbf {\bibinfo
  {volume} {57}},\ \bibinfo {pages} {3615} (\bibinfo {year}
  {1998})}\BibitemShut {NoStop}%
\bibitem [{\citenamefont {Morgan}\ and\ \citenamefont
  {Ketterson}(1998)}]{Morgan1998}%
  \BibitemOpen
  \bibfield  {author} {\bibinfo {author} {\bibfnamefont {D.~J.}\ \bibnamefont
  {Morgan}}\ and\ \bibinfo {author} {\bibfnamefont {J.~B.}\ \bibnamefont
  {Ketterson}},\ }\href {\doibase 10.1103/PhysRevLett.80.3614} {\bibfield
  {journal} {\bibinfo  {journal} {Phys. Rev. Lett.}\ }\textbf {\bibinfo
  {volume} {80}},\ \bibinfo {pages} {3614} (\bibinfo {year}
  {1998})}\BibitemShut {NoStop}%
\bibitem [{\citenamefont {Villegas}\ \emph {et~al.}(2003)\citenamefont
  {Villegas}, \citenamefont {Gonzalez}, \citenamefont {Montero}, \citenamefont
  {Schuller},\ and\ \citenamefont {Vicent}}]{Villegas2003}%
  \BibitemOpen
  \bibfield  {author} {\bibinfo {author} {\bibfnamefont {J.~E.}\ \bibnamefont
  {Villegas}}, \bibinfo {author} {\bibfnamefont {E.~M.}\ \bibnamefont
  {Gonzalez}}, \bibinfo {author} {\bibfnamefont {M.~I.}\ \bibnamefont
  {Montero}}, \bibinfo {author} {\bibfnamefont {{\relax Ivan K}.}~\bibnamefont
  {Schuller}}, \ and\ \bibinfo {author} {\bibfnamefont {J.~L.}\ \bibnamefont
  {Vicent}},\ }\href {\doibase 10.1103/PhysRevB.68.224504} {\bibfield
  {journal} {\bibinfo  {journal} {Phys. Rev. B}\ }\textbf {\bibinfo {volume}
  {68}},\ \bibinfo {pages} {224504} (\bibinfo {year} {2003})}\BibitemShut
  {NoStop}%
\bibitem [{\citenamefont {Kemmler}\ \emph {et~al.}(2006)\citenamefont
  {Kemmler}, \citenamefont {G\"urlich}, \citenamefont {Sterck}, \citenamefont
  {P\"ohler}, \citenamefont {Neuhaus}, \citenamefont {Siegel}, \citenamefont
  {Kleiner},\ and\ \citenamefont {Koelle}}]{Kemmler2006}%
  \BibitemOpen
  \bibfield  {author} {\bibinfo {author} {\bibfnamefont {M.}~\bibnamefont
  {Kemmler}}, \bibinfo {author} {\bibfnamefont {C.}~\bibnamefont {G\"urlich}},
  \bibinfo {author} {\bibfnamefont {A.}~\bibnamefont {Sterck}}, \bibinfo
  {author} {\bibfnamefont {H.}~\bibnamefont {P\"ohler}}, \bibinfo {author}
  {\bibfnamefont {M.}~\bibnamefont {Neuhaus}}, \bibinfo {author} {\bibfnamefont
  {M.}~\bibnamefont {Siegel}}, \bibinfo {author} {\bibfnamefont
  {R.}~\bibnamefont {Kleiner}}, \ and\ \bibinfo {author} {\bibfnamefont
  {D.}~\bibnamefont {Koelle}},\ }\href {\doibase 10.1103/PhysRevLett.97.147003}
  {\bibfield  {journal} {\bibinfo  {journal} {Phys. Rev. Lett.}\ }\textbf
  {\bibinfo {volume} {97}},\ \bibinfo {pages} {147003} (\bibinfo {year}
  {2006})}\BibitemShut {NoStop}%
\bibitem [{\citenamefont {Misko}\ \emph {et~al.}(2010)\citenamefont {Misko},
  \citenamefont {Bothner}, \citenamefont {Kemmler}, \citenamefont {Kleiner},
  \citenamefont {Koelle}, \citenamefont {Peeters},\ and\ \citenamefont
  {Nori}}]{Misko2010}%
  \BibitemOpen
  \bibfield  {author} {\bibinfo {author} {\bibfnamefont {V.~R.}\ \bibnamefont
  {Misko}}, \bibinfo {author} {\bibfnamefont {D.}~\bibnamefont {Bothner}},
  \bibinfo {author} {\bibfnamefont {M.}~\bibnamefont {Kemmler}}, \bibinfo
  {author} {\bibfnamefont {R.}~\bibnamefont {Kleiner}}, \bibinfo {author}
  {\bibfnamefont {D.}~\bibnamefont {Koelle}}, \bibinfo {author} {\bibfnamefont
  {F.~M.}\ \bibnamefont {Peeters}}, \ and\ \bibinfo {author} {\bibfnamefont
  {F.}~\bibnamefont {Nori}},\ }\href {\doibase 10.1103/PhysRevB.82.184512}
  {\bibfield  {journal} {\bibinfo  {journal} {Phys. Rev. B}\ }\textbf {\bibinfo
  {volume} {82}},\ \bibinfo {pages} {184512} (\bibinfo {year}
  {2010})}\BibitemShut {NoStop}%
\bibitem [{\citenamefont {Rosen}\ \emph {et~al.}(2010)\citenamefont {Rosen},
  \citenamefont {Sharoni},\ and\ \citenamefont {Schuller}}]{Rosen2010}%
  \BibitemOpen
  \bibfield  {author} {\bibinfo {author} {\bibfnamefont {Y.~J.}\ \bibnamefont
  {Rosen}}, \bibinfo {author} {\bibfnamefont {A.}~\bibnamefont {Sharoni}}, \
  and\ \bibinfo {author} {\bibfnamefont {{\relax Ivan K}.}~\bibnamefont
  {Schuller}},\ }\href {\doibase 10.1103/PhysRevB.82.014509} {\bibfield
  {journal} {\bibinfo  {journal} {Phys. Rev. B}\ }\textbf {\bibinfo {volume}
  {82}},\ \bibinfo {pages} {014509} (\bibinfo {year} {2010})}\BibitemShut
  {NoStop}%
\bibitem [{\citenamefont {Kemmler}\ \emph {et~al.}(2009)\citenamefont
  {Kemmler}, \citenamefont {Bothner}, \citenamefont {Ilin}, \citenamefont
  {Siegel}, \citenamefont {Kleiner},\ and\ \citenamefont
  {Koelle}}]{Kemmler2009}%
  \BibitemOpen
  \bibfield  {author} {\bibinfo {author} {\bibfnamefont {M.}~\bibnamefont
  {Kemmler}}, \bibinfo {author} {\bibfnamefont {D.}~\bibnamefont {Bothner}},
  \bibinfo {author} {\bibfnamefont {K.}~\bibnamefont {Ilin}}, \bibinfo {author}
  {\bibfnamefont {M.}~\bibnamefont {Siegel}}, \bibinfo {author} {\bibfnamefont
  {R.}~\bibnamefont {Kleiner}}, \ and\ \bibinfo {author} {\bibfnamefont
  {D.}~\bibnamefont {Koelle}},\ }\href {\doibase 10.1103/PhysRevB.79.184509}
  {\bibfield  {journal} {\bibinfo  {journal} {Phys. Rev. B}\ }\textbf {\bibinfo
  {volume} {79}},\ \bibinfo {pages} {184509} (\bibinfo {year}
  {2009})}\BibitemShut {NoStop}%
\bibitem [{\citenamefont {Misko}\ and\ \citenamefont {Nori}(2012)}]{Misko2012}%
  \BibitemOpen
  \bibfield  {author} {\bibinfo {author} {\bibfnamefont {V.~R.}\ \bibnamefont
  {Misko}}\ and\ \bibinfo {author} {\bibfnamefont {F.}~\bibnamefont {Nori}},\
  }\href {\doibase 10.1103/PhysRevB.85.184506} {\bibfield  {journal} {\bibinfo
  {journal} {Phys. Rev. B}\ }\textbf {\bibinfo {volume} {85}},\ \bibinfo
  {pages} {184506} (\bibinfo {year} {2012})}\BibitemShut {NoStop}%
\bibitem [{\citenamefont {Ray}\ \emph {et~al.}(2012)\citenamefont {Ray},
  \citenamefont {Reichhardt}, \citenamefont {Janko},\ and\ \citenamefont
  {Reichhardt}}]{Ray2012}%
  \BibitemOpen
  \bibfield  {author} {\bibinfo {author} {\bibfnamefont {D.}~\bibnamefont
  {Ray}}, \bibinfo {author} {\bibfnamefont {C.~J.~Olson}\ \bibnamefont
  {Reichhardt}}, \bibinfo {author} {\bibfnamefont {B.}~\bibnamefont {Janko}}, \
  and\ \bibinfo {author} {\bibfnamefont {C.}~\bibnamefont {Reichhardt}},\
  }\href@noop {} {\bibfield  {journal} {\bibinfo  {journal} {arXiv:1210.1229}\
  } (\bibinfo {year} {2012})}\BibitemShut {NoStop}%
\bibitem [{\citenamefont {Rothen}\ \emph {et~al.}(1993)\citenamefont {Rothen},
  \citenamefont {Pieranski}, \citenamefont {Rivier},\ and\ \citenamefont
  {Joyet}}]{Rothen1993}%
  \BibitemOpen
  \bibfield  {author} {\bibinfo {author} {\bibfnamefont {F.}~\bibnamefont
  {Rothen}}, \bibinfo {author} {\bibfnamefont {P.}~\bibnamefont {Pieranski}},
  \bibinfo {author} {\bibfnamefont {N.}~\bibnamefont {Rivier}}, \ and\ \bibinfo
  {author} {\bibfnamefont {A.}~\bibnamefont {Joyet}},\ }\href
  {http://stacks.iop.org/0143-0807/14/i=5/a=007} {\bibfield  {journal}
  {\bibinfo  {journal} {Eur. Phys. J.}\ }\textbf {\bibinfo {volume} {14}},\
  \bibinfo {pages} {227} (\bibinfo {year} {1993})}\BibitemShut {NoStop}%
\bibitem [{\citenamefont {Rothen}\ and\ \citenamefont
  {Pieranski}(1996)}]{Rothen1996}%
  \BibitemOpen
  \bibfield  {author} {\bibinfo {author} {\bibfnamefont {F.}~\bibnamefont
  {Rothen}}\ and\ \bibinfo {author} {\bibfnamefont {P.}~\bibnamefont
  {Pieranski}},\ }\href {\doibase 10.1103/PhysRevE.53.2828} {\bibfield
  {journal} {\bibinfo  {journal} {Phys. Rev. E}\ }\textbf {\bibinfo {volume}
  {53}},\ \bibinfo {pages} {2828} (\bibinfo {year} {1996})}\BibitemShut
  {NoStop}%
\bibitem [{\citenamefont {V\'elez}\ \emph {et~al.}(2008)\citenamefont
  {V\'elez}, \citenamefont {Mart\'in}, \citenamefont {Villegas}, \citenamefont
  {Hoffmann}, \citenamefont {Gonz\'alez}, \citenamefont {Vicent},\ and\
  \citenamefont {Schuller}}]{Velez2008}%
  \BibitemOpen
  \bibfield  {author} {\bibinfo {author} {\bibfnamefont {M.}~\bibnamefont
  {V\'elez}}, \bibinfo {author} {\bibfnamefont {J.~I.}\ \bibnamefont
  {Mart\'in}}, \bibinfo {author} {\bibfnamefont {J.~E.}\ \bibnamefont
  {Villegas}}, \bibinfo {author} {\bibfnamefont {A.}~\bibnamefont {Hoffmann}},
  \bibinfo {author} {\bibfnamefont {E.~M.}\ \bibnamefont {Gonz\'alez}},
  \bibinfo {author} {\bibfnamefont {J.~L.}\ \bibnamefont {Vicent}}, \ and\
  \bibinfo {author} {\bibfnamefont {{\relax Ivan K}.}~\bibnamefont
  {Schuller}},\ }\href {\doibase 10.1016/j.jmmm.2008.06.013} {\bibfield
  {journal} {\bibinfo  {journal} {J. Magn. Magn. Mater.}\ }\textbf {\bibinfo
  {volume} {320}},\ \bibinfo {pages} {2547 } (\bibinfo {year}
  {2008})}\BibitemShut {NoStop}%
\bibitem [{\citenamefont {Daire}\ \emph {et~al.}(2005)\citenamefont {Daire},
  \citenamefont {Goeke},\ and\ \citenamefont {Tupta}}]{Daire2005}%
  \BibitemOpen
  \bibfield  {author} {\bibinfo {author} {\bibfnamefont {A.}~\bibnamefont
  {Daire}}, \bibinfo {author} {\bibfnamefont {W.}~\bibnamefont {Goeke}}, \ and\
  \bibinfo {author} {\bibfnamefont {M.}~\bibnamefont {Tupta}},\ }\href@noop {}
  {\emph {\bibinfo {title} {White Paper:New Instruments Can Lock Out
  Lock-ins}}},\ \bibinfo {type} {Tech. Rep.}\ (\bibinfo  {institution}
  {Keithley Instruments, Inc., Cleveland, OH},\ \bibinfo {year}
  {2005})\BibitemShut {NoStop}%
\bibitem [{\citenamefont {Hoffmann}\ \emph {et~al.}(2000)\citenamefont
  {Hoffmann}, \citenamefont {Prieto},\ and\ \citenamefont
  {Schuller}}]{Hoffmann2000}%
  \BibitemOpen
  \bibfield  {author} {\bibinfo {author} {\bibfnamefont {A.}~\bibnamefont
  {Hoffmann}}, \bibinfo {author} {\bibfnamefont {P.}~\bibnamefont {Prieto}}, \
  and\ \bibinfo {author} {\bibfnamefont {{\relax Ivan K}.}~\bibnamefont
  {Schuller}},\ }\href {\doibase 10.1103/PhysRevB.61.6958} {\bibfield
  {journal} {\bibinfo  {journal} {Phys. Rev. B}\ }\textbf {\bibinfo {volume}
  {61}},\ \bibinfo {pages} {6958} (\bibinfo {year} {2000})}\BibitemShut
  {NoStop}%
\bibitem [{\citenamefont {W\"ordenweber}\ \emph {et~al.}(2000)\citenamefont
  {W\"ordenweber}, \citenamefont {Castellanos},\ and\ \citenamefont
  {Selders}}]{Woerdenweber2000}%
  \BibitemOpen
  \bibfield  {author} {\bibinfo {author} {\bibfnamefont {R.}~\bibnamefont
  {W\"ordenweber}}, \bibinfo {author} {\bibfnamefont {A.}~\bibnamefont
  {Castellanos}}, \ and\ \bibinfo {author} {\bibfnamefont {P.}~\bibnamefont
  {Selders}},\ }\href {\doibase 10.1016/S0921-4534(99)00639-5} {\bibfield
  {journal} {\bibinfo  {journal} {Physica C:}\ }\textbf {\bibinfo {volume}
  {332}},\ \bibinfo {pages} {27 } (\bibinfo {year} {2000})}\BibitemShut
  {NoStop}%
\bibitem [{\citenamefont {Bothner}\ \emph {et~al.}(2011)\citenamefont
  {Bothner}, \citenamefont {Gaber}, \citenamefont {Kemmler}, \citenamefont
  {Koelle},\ and\ \citenamefont {Kleiner}}]{Bothner2011}%
  \BibitemOpen
  \bibfield  {author} {\bibinfo {author} {\bibfnamefont {D.}~\bibnamefont
  {Bothner}}, \bibinfo {author} {\bibfnamefont {T.}~\bibnamefont {Gaber}},
  \bibinfo {author} {\bibfnamefont {M.}~\bibnamefont {Kemmler}}, \bibinfo
  {author} {\bibfnamefont {D.}~\bibnamefont {Koelle}}, \ and\ \bibinfo {author}
  {\bibfnamefont {R.}~\bibnamefont {Kleiner}},\ }\href {\doibase
  10.1063/1.3560480} {\bibfield  {journal} {\bibinfo  {journal} {Appl. Phys.
  Lett.}\ }\textbf {\bibinfo {volume} {98}},\ \bibinfo {eid} {102504} (\bibinfo
  {year} {2011})}\BibitemShut {NoStop}%
\end{thebibliography}
\end{document}